%% file: main.tex
\documentclass[%
reprint,
superscriptaddress,
amsmath,amssymb,
aps,
prl,
floatfix,
longbibliography
]{revtex4-2}

\usepackage{times}
\usepackage{amstext}
\usepackage{graphicx}
\usepackage{color}
\usepackage{soul}
\usepackage{MnSymbol}
\usepackage{subfiles}
\usepackage{bm}
\usepackage{hyperref}
\usepackage{layouts}
\usepackage{enumitem}
\usepackage{natbib}
\usepackage{comment}
\usepackage{xcolor}

\newcommand{\dd}{\mathrm{d}}
\newcommand{\e}{\mathrm{e}}

\newcommand{\sign}{\operatorname{sign}}

\newcommand{\rhoI}{\rho_{\mathrm I}}
\newcommand{\rhoA}{\rho_{\mathrm A}}

\begin{document}

\title{Hydrodynamics of Cooperation and Self-Interest in a Two-Population Occupation Model}

\author{Jérôme Garnier-Brun}
\thanks{Both authors contributed equally, alphabetical order.}
\affiliation{Department of Computing Sciences \& Department of Finance, Università Bocconi, Milan, Italy}
\affiliation{Chair of Econophysics and Complex Systems, École Polytechnique, 91128 Palaiseau Cedex, France}

\author{Ruben Zakine}
\thanks{Both authors contributed equally, alphabetical order.}
\affiliation{Chair of Econophysics and Complex Systems, École Polytechnique, 91128 Palaiseau Cedex, France}
\affiliation{LadHyX UMR CNRS 7646, École Polytechnique, 91128 Palaiseau Cedex, France}

\author{Michael Benzaquen}
\affiliation{Chair of Econophysics and Complex Systems, École Polytechnique, 91128 Palaiseau Cedex, France}
\affiliation{LadHyX UMR CNRS 7646, École Polytechnique, 91128 Palaiseau Cedex, France}
\affiliation{Capital Fund Management, 23 Rue de l’Université, 75007 Paris, France}

\begin{abstract}
We study the hydrodynamics of a system of agents who optimize either their individual utility (self-interest) or the collective welfare (cooperation). When agents act selfishly, their interactions are non-reciprocal, driving the system out of equilibrium; by contrast, purely altruistic dynamics restore reciprocity and yield an equilibrium-like description. We investigate how mixtures of these two behaviors shape the macroscopic properties of the liquid of agents. For highly rational agents, we find that introducing a small fraction of altruists can suppress the sub-optimal clustering induced by selfish dynamics. This phenomenon can be attributed to altruists localizing at interfaces and acting as effective surfactants, shedding a new light on earlier findings in fixed neighborhood-based models [Phys. Rev. Lett. \textbf{120}, 208301 (2018)].
When agents are boundedly rational, we introduce a well-mixed approximation that reduces the two-population model to a single effective scalar field theory. This allows us to leverage state-of-the-art tools from active matter to analytically characterize how altruism modifies surface tension and nucleation dynamics.

\end{abstract}

\date{\today}

\maketitle

Equilibrium statistical mechanics describes the collective behavior of a large number of constituents when their dynamics are driven by the minimization of a globally defined energy. In many complex systems, however, finding such a system-wide objective function may be impossible. 
This is notably the case e.g. in active matter~\cite{ramaswamy2010mechanics,marchetti2013hydrodynamics,stenhammar2013continuum,wittkowski2014scalar}, where particles locally inject energy and momentum in the medium, or else in {biological} neural networks~\cite{hertz1987irreversible,marti2018correlations,corral2022excitatory}, in which non-reciprocal interactions are commonly assumed. Another wide class of problems where finding a global quantity to be minimized is challenging, if not impossible, is socioeconomic systems. 
Indeed, most often one cannot  assume that individuals share a common ``utility" they all strive to optimize; it seems more realistic to consider agents as individualistic actors seeking to improve their own satisfaction, possibly at the expense of the wider population, as pinpointed by the pioneering agent-based models by Sakoda~\cite{sakoda_checkerboard_1971} and Schelling~\cite{Schelling1971} (see e.g.~\cite{bouchaud2013crises,garnier2023bounded} for discussions on this issue). In this context, ~\citet{zakine2024socioeconomic} notably showed that in an occupation model where a fixed number of individualistic agents populate a lattice depending on their \textit{own} preference, detailed balance is violated at the ``microscopic'' level, and this regardless of the details of the agent's decision rules. Coarse-graining the system, the density of agents follows a stochastic hydrodynamic equation in which the driving term cannot be written as a gradient of a free energy functional, placing the system out of equilibrium~\cite{nardini2017entropy,o2023nonequilibrium}.

Importantly, when such socioeconomic systems reach a nonequilibrium steady state due to individualistic decision-making, the collective welfare—captured by the average utility of agents—remains a meaningful observable that can possibly be measured by surveys~\cite{seara2023sociohydrodynamics}. Assessing the penalty due to individualistic behavior on this global utility is therefore a key question in social dynamics, and motivates bridging between nonequilibrium and equilibrium (i.e. \textit{global} utility-maximizing) descriptions.
Exploring this interplay between local and global optimization is particularly relevant at the hydrodynamic level, where coarse-grained descriptions provide a natural framework for integrating empirical socioeconomic data, as recently demonstrated by Seara \textit{et al.}~\cite{seara2023sociohydrodynamics}. 
Beyond the socioeconomic context, this competition between local and global optimization logics is also likely to play a key role in the emerging field of decentralized learning~\cite{long2018towards,bredeche2022social,ben2023morphological}, where recent studies~\cite{jung2025kinetic} have begun to explore hydrodynamic descriptions of learning agents optimizing individual rewards---i.e. ``smart'' active matter.

In this Letter, we address these questions by considering the interaction between individualistic and altruistic agents, who maximize the system-wide aggregate utility instead of their own. By placing ourselves at the level of nonequilibrium hydrodynamics, where the coarse-grained properties of the system are effectively captured, our model extends the results of~\citet{grauwin2009competition} and~\citet{jensen2018giant} to a more general and physically interpretable framework (see below). 
Having uncovered the salient effect of altruism in our two-population setting, we then introduce a ``well-mixed'' approximation of our model, with a single population making their decision based either on their personal satisfaction or on the collective well-being with a given probability. This reduced setting allows us to leverage recent advances in the theory of active matter, namely the so-called generalized thermodynamic mapping~\cite{solon2018generalized1,solon2018generalized2}, thereby providing a physical interpretation of the effect of altruism through the change of properties of the ``liquid'' of agents. 

\paragraph*{Model.} Suppose an individual is endowed with a utility function $u$. This utility function is a measure of their satisfaction at a position $x \in [0,L]^d$ in, say, a city, and reads
\begin{equation}
    u(\phi(x)) = - \lvert\phi(x) - \rho^\star\rvert^\gamma, \qquad \gamma > 0
    \label{eq:individual_utility}
\end{equation}
 where $\phi(x)$ is a locally perceived density of neighbors, and $0 < \rho^\star < 1$ is the community-wide ideal of surrounding occupation density~\footnote{{Henceforth, we shall always assume periodic boundary conditions and arbitrarily set $\gamma = 3/2$.}}. The perceived density is given by $\phi(x) \equiv \phi([\rho],x) = (G * \rho )(x)$, 
where the density kernel $G$ of standard deviation $\sigma$ is generically expected to  decrease monotonically, while $\rho(x)$ is simply the population density field in this idealized world. As the utility function $u(z)$ described in Eq.~\eqref{eq:individual_utility} is non-monotonic, our toy model essentially relies on the following assumption: Individuals wish to reside in an area that is not too empty, as they want to enjoy a rich social environment and have access to a number of services, but that is not too full either, in order to benefit from a good quality of life and high level of comfort.

As mentioned above, a common starting point for such a model is to assume that agents behave such as to maximize their own satisfaction, that is the utility function evaluated at the position they choose to live in. 
Then, imagine some purely \textit{altruistic} agents who have the common interest in mind when making their decision. Instead of attempting at maximizing their own utility, these agents seek to improve the average outcome of the society, in our case proportional to the \textit{global} utility $\mathcal{U}[\rho] = \int \dd x \, u(\phi([\rho],x)) \rho(x)$, 
potentially at the cost of their own satisfaction \footnote{If such infinitely benevolent individuals may appear somewhat unrealistic, this behavior could also be seen to arise through the action of a central planner that would have the power to influence the decision of individuals, for instance through social housing.}. Space is occupied with a conserved global density $\rho_0$ of agents, split between {fixed fractions $\alpha$ of altruistic agents, and $1-\alpha$ of individualistic agents}. This can be modeled with two coexisting and interacting density fields $\rho_{\mathrm A}(x,t)$ and $\rho_{\mathrm I}(x,t)$, describing the spatial and temporal distribution of altruists and individualists, respectively. Agents are blind to the ``type'' of other agents, e.g. an individualist perceives the presence of another individualist as equivalent to that of an altruist.

Let us start by writing the dynamics followed by the density of individualistic agents. The hydrodynamic equation can be derived from the ``microscopic'' (local) dynamics of non-overlapping agents on a $d$-dimensional lattice using a path integral approach~\cite{lefevre2007dynamics}, which is exact in the thermodynamic limit \footnote{To be precise, the thermodynamic limit here corresponds to infinitely many agents and vanishingly small lattice spacings.} (see Supplemental Material
), yielding
\begin{equation}
    \partial_t \rho_{\mathrm I} = \nabla \cdot \left( M_\mathrm{I}[\rhoI, \rho_A]\nabla \mu_{\mathrm I}[\rho_{\mathrm I},\rho_{\mathrm A}] + \sqrt{2T M_\mathrm{I}[\rhoI,\rhoA]}\xi_\mathrm{I}\right),
    \label{eq:model1_individualist}
\end{equation}
with $M_\mathrm{I}[\rhoI,\rhoA]= \rhoI(1-\rhoA-\rhoI)$ a standard non-overlapping motility, $\xi_\mathrm{I}$ a Gaussian white noise in space and time, and $\mu_{\mathrm I}$ an effective chemical potential,
\begin{equation}
    \mu_{\mathrm I}[\rho_{\mathrm I},\rho_{\mathrm A}] = T \log\left(\frac{\rho_{\mathrm I}}{1-\rho_{\mathrm I} - \rho_{\mathrm A}} \right) -  u(\phi([\rho_{\mathrm I}+\rho_{\mathrm A}],x)).
\end{equation}
The first term is an entropic contribution (i.e. diffusion of the density field), with $T$ a temperature parametrizing the fluctuations in the decision-making process of the agents. This is rather standard in the socioeconomic literature~\cite{bouchaud2013crises,anderson1992discrete}, where the inverse temperature is referred to as the ``rationality'', or more precisely as the ``intensity of choice''. The second term can be understood as a consequence of agents maximizing their utility (cleverly coined ``utility-taxis'' in~\cite{seara2023sociohydrodynamics}), consistent with the link between utility and chemical potential first proposed in \cite{lemoy2011socio}. Importantly, we showed in~\cite{zakine2024socioeconomic} that a utility function that is nonlinear in the \textit{perceived} density $\phi$ \textit{cannot} be written as the functional derivative of a global functional of the density field. As a result, the steady-state solution of Eq.~\eqref{eq:model1_individualist} may not be described with the standard tools of equilibrium statistical mechanics. Note that this significantly differs from Refs.~\cite{grauwin2009competition} and~\cite{jensen2018giant}. Indeed, both these seminal works leveraged a predefined district-based construction that singularly allows a system of individualists to minimize an effective free energy. Yet, the existence of such an object in socioeconomic systems is the exception rather than the rule~\cite{bouchaud2013crises}; assessing the robustness of results to an inherently nonequilibrium setting, particularly at the coarse-grained level, is therefore an important question.

Altruists, on the other hand, follow the dynamics
\begin{equation}
    \partial_t \rho_{\mathrm A} = \nabla\cdot \left(M_\mathrm{A}[\rhoA,\rhoI] \nabla \frac{\delta \mathcal F }{\delta \rho_{\mathrm A}} + \sqrt{2T M_\mathrm{A}[\rhoA,\rhoI]} \xi_\mathrm{A}\right),
    \label{eq:model1_altruist}
\end{equation}
with $M_\mathrm{A}[\rhoA,\rhoI]=\rhoA(1-\rhoA-\rhoI)$, $\xi_\mathrm{A}$ a Gaussian white noise, and where now $\mathcal{F}[\rho_{\mathrm I},\rho_{\mathrm A}] = -\mathcal{U}[\rho_{\mathrm I} + \rho_{\mathrm A}] - T \mathcal{S}[\rho_{\mathrm I},\rho_{\mathrm A}]$ is a standard free energy functional, sum of the (minus) global utility, and of the entropy of mixing
$\mathcal{S}[\rho_{\mathrm I},\rho_{\mathrm A}] = -\int  \{\rho_{\mathrm I} \log\rho_{\mathrm I} + \rho_{\mathrm A}\log\rho_{\mathrm A} + (1-\rho_{\mathrm A}-\rho_{\mathrm I})\log\left(1-\rho_{\mathrm A}-\rho_{\mathrm I} \right)\}$. 
The coupled Eqs.~\eqref{eq:model1_individualist} and \eqref{eq:model1_altruist}, together with a prescribed initial density field, describe our \textit{non-reciprocal} hydrodynamic model \footnote{Note that the coupled hydrodynamic model is necessarily out of equilibrium due to the inherently non-reciprocal nature of the coupling between the two types of agents, and this even if the individualistic chemical potential can be written as the functional derivative of a free energy functional, as pointed out in~\cite{jensen2018giant} where the $\alpha = 0$ case is in equilibrium.}. When {the altruistic fraction $\alpha = 1$ (no individualists)}, the above equation corresponds to equilibrium dynamics. 
Note finally that a hypothetical central planner is interested in the global utility $\mathcal U$ describing the system, irrespective of the agents being individualist or  altruist, and  the system being in or out of equilibrium. 

\begin{figure}
    \centering
    \includegraphics[width=\linewidth]{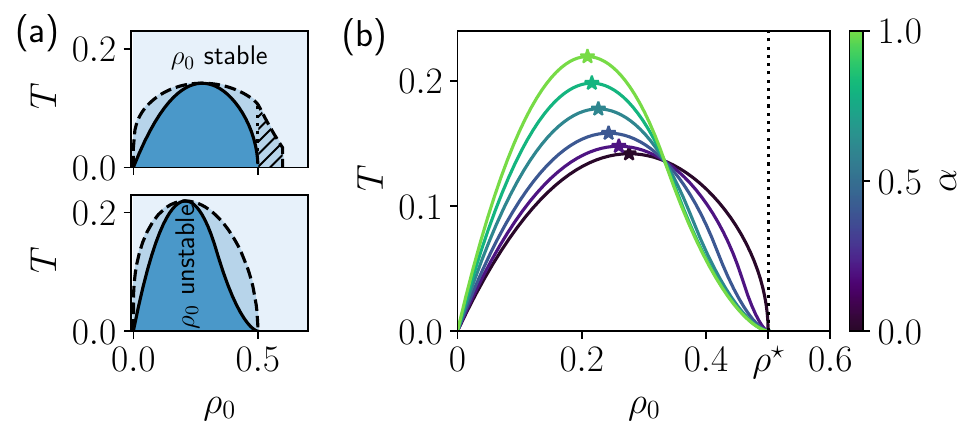}
    \caption{(a) Phase diagrams in the $\alpha = 0$ (top) and $\alpha = 1$ (bottom) cases with $\gamma = 3/2$ and $\rho^\star = 1/2$, full lines indicating the spinodals, and dashed lines delimiting the binodal region where the homogeneous state is metastable. In the hatched area, concentration is possible despite $\rho_0 \geq \rho^\star$. (b) Spinodal curves for different $\alpha$.}
    \label{fig:spinodals}
\end{figure} 

\paragraph*{Impact of altruism.} Before getting into the description of the model when $\alpha > 0$, we summarize the phenomenology of the $\alpha = 0$ field theory describing purely individualistic agents, as detailed in~\cite{zakine2024socioeconomic}. For small temperatures (high rationality), agents aggregate in dense suboptimal clusters. At the mean-field level, that is neglecting the noise in the hydrodynamic equation, the  spinodal and binodal curves can be determined analytically, resulting in the phase diagram shown in Fig.~\ref{fig:spinodals}(a), top panel \footnote{Note that when the smoothing kernel $G$ has a sufficiently large range, the noiseless approximation yields excellent predictions for this model~\cite{zakine2024socioeconomic}.}. Strikingly, the binodal curve extends to $\rho_0 > \rho^\star$: Due to the individualistic nature of the agents, coordination fails and the system ends up in significantly sub-optimal concentrated states even when a homogeneous distribution would provide a higher utility on average. We now seek to determine when and how the introduction of altruism affects this possibly sub-optimal collective behavior.

We first consider the linear stability of the homogeneous state $\rho_\mathrm{A}(x,t) = \alpha \rho_0$ and $\rho_\mathrm{I}(x,t) = (1-\alpha) \rho_0$ in our model. Introducing a small perturbation about this homogeneous state, expanding Eqs.~\eqref{eq:model1_individualist} and \eqref{eq:model1_altruist} at leading order and going to Fourier space yields a stability matrix, the eigenvalues of which can be computed analytically, see Supplemental Material (SM). Spinodal curves for $\alpha>0$ are shown in Fig.~\ref{fig:spinodals}(b). A first important observation is that in the vicinity of $\rho_0 = \rho^\star = 1/2$, a reasonable fraction of altruists leads the spinodal to lie very close to the equilibrium $\alpha = 1$ curve, leading to quasi-overlapping spinodals for $\alpha  \gtrapprox 1/2$. 
For smaller values of $\rho_0$, we observe a significant rise in the temperature of the critical point, corresponding to the maximum of the spinodal curves, as $\alpha$ is increased. In other words, the introduction of altruism allows agents to coordinate in a way that is more robust to random fluctuations. Below the critical temperature, we expect a metastable region, delimited by binodal curves, which we must now determine in order to properly describe the phase separation of the system.

\begin{figure}
    \centering
    \includegraphics[width=\linewidth]{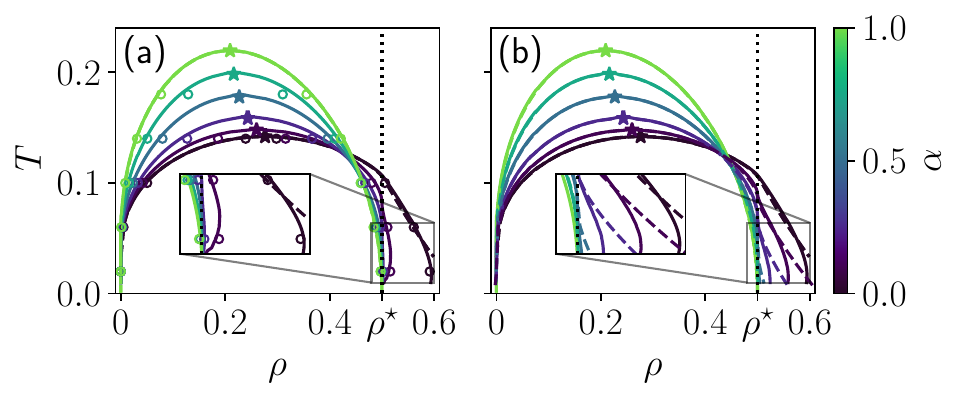}
    \caption{(a) Binodal curves ($\rho = \rhoA + \rhoI$); markers ($\circ$) show numerical estimates of the coexistence densities from agent-based simulations in $d=2$ (SM). (b) Binodal curves in the  case of an effective single-agent population. Dashed lines show the  analytic computation using the free energy for $\alpha = 1$ and the generalized thermodynamics approximation (only available for $\alpha = 0$ in (a)). Solid lines are constructed from the numerical resolution of the noiseless hydrodynamic PDEs with Gaussian $G$, $\sigma = 10$, $L = 1000$.}
    \label{fig:binodals}
\end{figure}

In the limiting case  $\alpha = 1$ (fully altruistic), corresponding to equilibrium dynamics, the binodal curve can be straightforwardly determined from the free energy density. Within the bulk of the assumed ``liquid'' (dense) and ``gas'' (close to empty) phases of the system, the local free energy density is $f(\rho) = -\rho u(\rho) + T \left[ \rho \log \rho + (1-\rho)\log(1-\rho)\right]$.
Performing the double-tangent construction on this function yields {the liquid-gas coexistence densities delimiting the binodal, $\rho_\ell$ and $\rho_g$ respectively,} for a given couple $(\rho_0,T)$. The resulting phase diagram in this $\alpha = 1$ case is shown in Fig.~\ref{fig:spinodals}(a), bottom panel. Importantly, one can see that it does not go beyond $\rho = \rho^\star$, meaning that, as expected, the system no longer sub-optimally phase separates when a uniform distribution of agents is preferable. 

What happens for intermediate values of $\alpha$? While the coupled nature of the dynamics prevents us from answering this question analytically, a simple alternative is to set the initial total density $\rho_0$ to its critical value obtained from linear stability analysis, and to numerically solve the system of noiseless partial differential equations (PDE) while varying temperature. The densities measured in the bulk of the liquid and gas phases then give a numerical estimate of the two branches of the binodal, which we show in Fig.~\ref{fig:binodals}(a). We verify that in the $\alpha = 1$ case, these densities perfectly match the equilibrium theoretical result. In addition to this numerical solution at the coarse-grained level, we have also performed agent-based numerical simulations on two-dimensional  lattices, see SM for details. The coexistence densities measured from these simulations are shown by the markers on Fig.~\ref{fig:binodals}(a), displaying an excellent match with the mean-field predictions. In the thermodynamic limit, the coexistence densities described by the binodals can be explicitly related to the global utility of the system, $\frac{\mathcal{U}}{L^d} = \frac{\rho_0 - \rho_g}{\rho_\ell - \rho_g} u(\rho_\ell) + \frac{\rho_\ell - \rho_0}{\rho_\ell - \rho_g} u(\rho_g) + O\left(\frac1L\right)$.
This relation allows us to directly transpose the consequences of altruism described below in terms of agents' welfare.

\begin{figure*}
    \centering
    \includegraphics[width=0.93\linewidth]{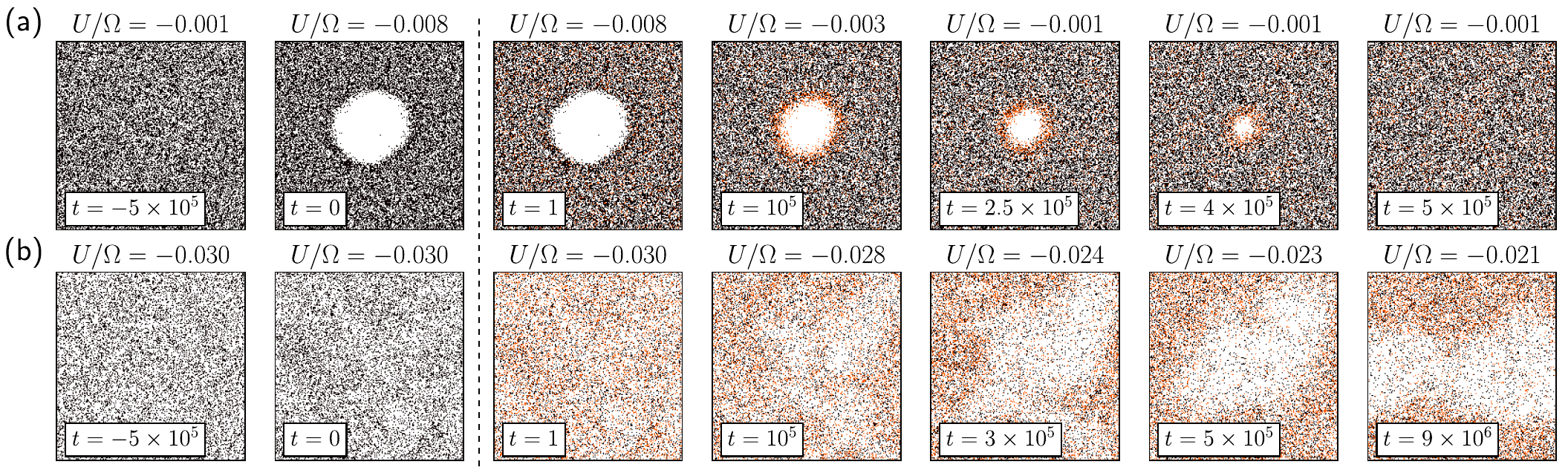}
    \caption{Illustration of the effects of altruists in an agent-based simulation of our model, see SM. In a system of size  $\Omega = L \times L$, with $L = 200$ {(periodic boundary conditions)}, of $N = \lfloor \rho_0 L^2 \rfloor$ individualists (black) are left to evolve from an initially homogeneous configuration. After $5\times 10^5$ Monte Carlo steps (vertical dashed line), a randomly selected fraction $\alpha$ individualists are replaced with altruists (red). (a)  $\rho_0 = \rho^\star = 1/2$, $T = 0.04$, $\alpha = 0.12$ (phase separation is unfavorable); (b)  $\rho_0 = 0.26$, $T = 0.14$, $\alpha = 0.5$ (phase separation is favorable). For both,  $G$ is Gaussian with $\sigma = 7$, $\gamma = 3/2$ and $\rho^\star = 1/2$.}
    \label{fig:bubble}
\end{figure*}

Altruism appears to have two markedly distinct effects on the binodals of the system. (\textit{i}) At low temperatures, a minute fraction of altruistic agents has a very strong effect on the aggregate behavior {and thus welfare} of the system. As shown in the inset of Fig.~\ref{fig:binodals}(a), small values of $\alpha$ indeed lead the agents to collectively behave as if the global utility was the quantity optimized by all agents, and almost immediately kill the sub-optimal concentration at densities $\rho > \rho^\star$ of the fully individualistic population. This very strong effect when $T$ is small, which appears to be compatible with the ``catalytic'' effect of altruism described in  Ref.~\cite{jensen2018giant} in a different setting, can be understood by observing the spatial distribution of the two types of agents. As visible in the agent-based simulation shown in Fig.~\ref{fig:bubble}(a), altruistic agents can very effectively inhibit the concentration of individualists by placing themselves at the boundary of population clusters, triggering a progressive spreading of the dense regions and effectively acting as surfactants. (\textit{ii})~At higher temperatures, consistent with the critical temperature computed above (Fig.~\ref{fig:spinodals}(b)), altruism allows the system to be significantly more robust to fluctuations, and to remain phase separated when it is beneficial for the agents. In this region, as shown in Fig.~\ref{fig:bubble}(b), the effect of $\alpha$ is not particularly localized in space. 

\paragraph*{Well-mixed approximation.}
The two-agent model discussed above demonstrates the strong impact of altruism. While we were able to unravel the mechanisms at low temperatures, we failed at describing it analytically beyond linear stability, and at clearly interpreting its effect at higher temperatures near the critical point. In order to improve our theoretical understanding, let us introduce an alternative version of the model with a single type of agents that happen to be altruistic at times. We now take $\alpha$ to be the probability of being altruistic at a given instant, leaving $1-\alpha$ probability to be individualistic (which is in fact similar to the prescription first proposed in~\cite{grauwin2009competition}). The resulting hydrodynamic equation for the single density field $\rho$
\begin{equation}
    \partial_t \rho = \nabla \cdot \left[ M[\rho] \nabla \mu_\mathrm{wm}[\rho] + \sqrt{2TM[\rho]} \xi\right], 
    \label{eq:model2}
\end{equation}
with $M[\rho]=\rho(1-\rho)$, $\mu_\mathrm{wm}[\rho]=(1-\alpha)\mu_\mathrm{I}[\rho_I=\rho,\rho_\mathrm{A}=0]+ \alpha\delta \mathcal F/\delta \rho$ that interpolates between individualist and altruist chemical potentials, $\mathcal F[\rho]=-\mathcal U[\rho]+T\int[\rho\log\rho+(1-\rho)\log(1-\rho)]$.
By design, this simplified model coincides with the original two-agent version in the extremal $\alpha = 0$ and $\alpha = 1$ cases.  In fact, both prescriptions are equivalent provided these two populations are well mixed in space, that is assuming $\frac{\rhoA(x)}{\rhoI(x)} = \frac{\alpha}{1-\alpha} \, \forall x$, which appears to be the case at sufficiently high temperature, see Fig.~\ref{fig:bubble}(b). By definition, this condition is verified in the homogeneous state where the density is uniform. Therefore both models have the same critical point and spinodal curves for a given value of $\alpha$, see SM.

The binodal curves in this setting can again be computed by numerically solving the PDE of Eq.~\eqref{eq:model2}. The resulting coexistence densities are shown as solid lines in Fig.~\ref{fig:binodals}(b). In the higher temperature region, the correspondence between the two models is quite remarkable, confirming that this simplified description may serve its purpose for the understanding of the role of altruism in the vicinity of $T_c(\alpha)$. For small $T$, the outcomes of the two prescriptions differ, {as can be expected from the spatially localized action of altruistic agents that breaks down the ``well-mixed'' assumption}, see  Fig.~\ref{fig:binodals} and Fig.~\ref{fig:bubble}(a).

Now, having a single scalar density field~$\rho$ is very convenient as it allows us to employ a generalized thermodynamics construction~\cite{solon2018generalized1,solon2018generalized2} after performing a gradient expansion of the chemical potential in Eq.~\eqref{eq:model2}:  $\mu_\mathrm{wm}([\rho],x) = \mu_0(\rho) + \lambda(\rho) (\nabla \rho)^2 - \kappa(\rho) \nabla^2 \rho + O(\nabla^4)$.
Taking the kernel $G$ to be Gaussian of range $\sigma$ yields explicit expressions for $\mu_0(\rho)$, $\kappa(\rho)$ and $\lambda(\rho)$, see SM.
The gradient expansion then suggests a bijective change of variable $R(\rho)$ with $\kappa(\rho)R''(\rho)=-[\kappa'(\rho)+2\lambda(\rho)]R'(\rho)$~\cite{solon2018generalized1,solon2018generalized2}, which restores locality in the phase properties and allows one to perform a double-tangent construction on a new generalized free energy density $g(R)$, defined such that $\mu_0(R) = \frac{\dd g}{\dd R}$.
Using Eq.~\eqref{eq:individual_utility} yields a (lengthy) analytical expression for $\mu_0(R)$, see SM. The predicted binodal densities are shown with dashed lines in Fig.~\ref{fig:binodals}(b). In the region of interest (upper part of the binodals), the match between the generalized thermodynamics analytical results and the numerically solved PDEs is excellent \footnote{At low temperatures, where the well-mixed approximation no longer holds,   the generalized thermodynamic mapping breaks down. This was already observed in the $\alpha = 0$ instance studied in~\cite{zakine2024socioeconomic}, and it can be understood from the gradient expansion. Indeed, considering the expression of $\kappa(\rho)$ that regulates interface stability to leading order, the non-monotonicity of the utility function leads to a change of sign of $\kappa$ at $\rho = \rho^\star(\alpha+1)/(1-\alpha+2\alpha\gamma)$, at which point one would require keeping higher order terms to stabilize the liquid-gas interface.}.

\begin{figure}
    \centering
    \includegraphics[width=\linewidth]{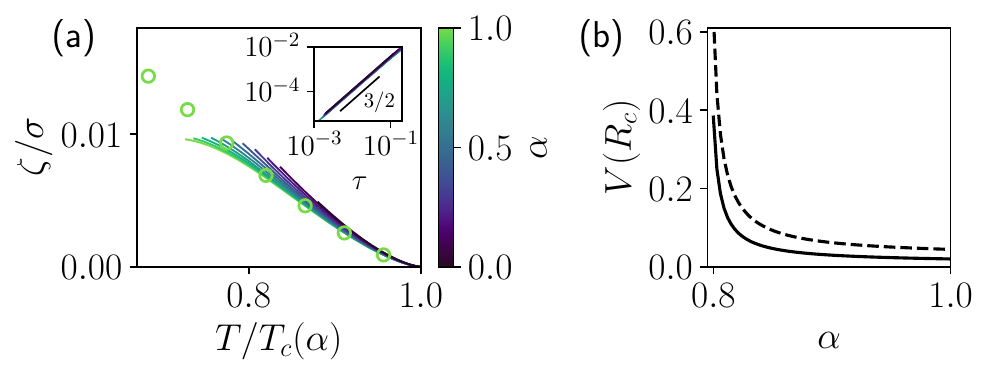}
    \caption{(a) Normalized pseudo tension $\zeta/\sigma$ (solid lines) as a function of rescaled temperature $T/T_c(\alpha)$, for different $\alpha$. Markers ($\circ$) indicate the true surface free energy in the equilibrium case $\alpha=1$. Inset: the tension follows the mean-field critical scaling $\zeta\propto \tau^s$ with $\tau=1-T/T_c(\alpha)$ and $s=3/2$~\cite{widom_surface_1965,Brilliantov2020}. (b)~Value of the quasi-potential $V(R_c)$ computed for fixed $\rho_0$, $T=0.18$ and close to the $\alpha = 0.8$ binodal. Solid line: close to the gas density $\rho_0=\rho_g(T)+\epsilon$. Dashed line: close to the liquid density $\rho_0=\rho_\ell(T)-\epsilon$, with $\epsilon=10^{-3}$.}
    \label{fig:tension}
\end{figure}

\paragraph*{Surface tension and nucleation.} 
Having verified the adequacy of the well-mixed approximation and of the gradient expansion close to the critical temperature, we leverage this analytically tractable setting to improve our understanding of the influence of altruism in this region. 
We notably propose to quantify the typical waiting times before observing a phase change as altruism is varied. Indeed, in the binodal regions the nucleation rate, which governs the time required for the agents to trigger phase separation, is ultimately $\alpha$ dependent.

The nucleation probability $\mathbb P$ (or nucleation rate) in active fluids was recently shown to follow classical nucleation theory~\cite{cates2023classical}. It satisfies a large deviation principle $\log \mathbb P \sim -V(R_c)/T$ for small $T$, where the quasi-potential $V(R_c)$ represents the cost to escape the homogeneous state and reach a critical nucleus {(a bubble of gas or a droplet of liquid)} of radius $R_c$, which then drives the system to complete phase separation. The quasi-potential $V(R_c)$ crucially depends on the surface tension, on the gas and liquid densities, and on the saturation (i.e. where we lie in the binodal). The surface tension of nonequilibrium liquids is a versatile quantity that should be interpreted with care~\cite{fausti_tjhung2021, zhao2024, langford2024}. This being said, the pseudo-tension obtained from the thermodynamic mapping~\cite{solon2018generalized1, solon2018generalized2} has been shown to regulate Ostwald ripening and therefore the fate of nucleation events in scalar active field theories~\cite{cates2023classical,tjhung_nardini2018}. 

The pseudo-tension $\zeta$ we compute from the gradient expansion (see SM) is shown for various $\alpha$ in Fig.~\ref{fig:tension}(a). For $\alpha=1$, we verify that the tension matches the true interfacial energy measured on a stationary profile close to the critical point. With this, one can finally compute the {quasi-}potential $V(R_c)$ and its dependence on $\alpha$. For a fixed $T$ \footnote{The choice of the temperature requires finding a ``sweet spot'' where $T$ is large enough for the pseudo surface tension $\zeta$ to be demonstrably accurate---that is not too far from $T_c(\alpha)$ (Fig.~\ref{fig:tension}(a))---, while being small enough for the large deviation principle of classical nucleation theory to remain valid. As we are mainly interested in the qualitative evolution of $V(R_c)$ with $\alpha$, we consider $T = 0.18$ to be adequate here.} and starting close to the binodal densities for, say, $\alpha=0.8$, one observes a sharp decrease of the quasi-potential as $\alpha$ increases, see Fig.~\ref{fig:tension}(b). Further, since the quasi-potential becomes comparable to $T$ as $\alpha$ increases, the nucleation rate is of order 1, and nucleation can thus no longer be seen as a rare event. These results are in line with the intuition that altruism radically facilitates the nucleation of the phase-separated state when it is beneficial for the agents, see Fig.~\ref{fig:bubble}(b).

\paragraph*{Concluding remarks.} 
In this work, we have considered a two-population extension of a Sakoda-Schelling occupation model to study the interaction between individualistic and altruistic agents and its nonlinear impact on global welfare at the hydrodynamic level. At low temperature, we find that a very small fraction of altruistic agents, effectively acting as surfactants, is highly effective at eliminating the sub-optimal clustering driven by selfish behavior. At higher temperatures, altruism drives the system in the phase-separated configuration, which here optimizes the global utility. In this region, the influence of altruism can be further understood through the lens of the probability of nucleation when adopting a single-population version of the model.

Let us finally discuss further the single-population prescription, as it can in fact be considered as a different model in itself. While it behaves similarly to the two-population case near criticality, it differs markedly at low temperature: Dedicated altruistic agents are far more effective than having the same fraction of decisions altruistically motivated at random. This highlights the importance of spatial localization, as altruists can concentrate at interfaces and progressively restore optimal configurations. Although derived in a simplified setting, we believe that this mechanism may be generic, with potential applications beyond socioeconomic systems---for instance, in the design of decentralized learning strategies for intelligent active matter, where balancing individual and collective optimization is key to controlling emergent behavior.

\paragraph*{Acknowledgments.}
We thank J.-P. Bouchaud for fruitful discussions.   This research was conducted within the Econophysics $\&$ Complex Systems Research Chair, under the aegis of the Fondation du Risque, the Fondation de l’École polytechnique, the École polytechnique and Capital Fund Management. JGB's research was developed  within the MUSA – Multilayered Urban Sustainability Action – project CUP B43D21011010006, funded by the European Union – NextGenerationEU, under the National Recovery and Resilience Plan (NRRP) Mission 4 Component 2 Investment Line 1.5: Strenghtening of research structures and creation of R\&D ``innovation ecosystems'', set up of ``territorial leaders in R\&D''.

\bibliography{refs}

\include{SM}

\end{document}

%% file: SM.tex
\onecolumngrid

\renewcommand{\theequation}{S\arabic{equation}}
\renewcommand{\thetable}{S\arabic{table}}
\renewcommand{\thefigure}{S\arabic{figure}}
\setcounter{equation}{0}
\setcounter{table}{0}
\setcounter{figure}{0}

\renewcommand{\thesection}{\Roman{section}}

\section{Supplemental Material}

{
\subsection{Derivation of the stochastic field theory from discrete agents}
Here, we follow the approach of \citet{lefevre2007dynamics} (see also \citet{andreanov2006field}), which was first detailed for the case of individualistic agents in \cite{garnier2023navigating}. We start by writing the generating function for particles locally diffusing on a lattice as
\begin{equation}
    Z[\{n,\hat{n}\}] = \int \{\dd n \, \dd \hat{n} \}\, \e^{-S[\{n,\hat{n}\}]},
\end{equation}
with the Martin-Siggia-Rose-Jansen-de Dominicis (MSRJD) action
\begin{equation}
    S[\{n,\hat{n}\}] = -\int \dd t \, \left\{ -\sum_i \hat{n}_i \partial_t n_i + \sum_{(i,j)} n_i W_{ij}(\e^{\hat{n}_j - \hat{n}_i} - 1) \right\},
    \label{eq:generating_func_discrete}
\end{equation}
where $W_{ij}$ is the transition rate from site $i$ to $j$. Note that here we consider the occupation variables as continuous. In the case of an exclusion process such as the one we are considering here (there can be no more than one agent at a given point in space), we take
\begin{equation}
    n_i W_{ij} = n_i (1-n_j) f_{ij},
\end{equation}
such that it takes the value 0 if site $i$ is empty and/or if site $j$ is occupied. Now, we assume that the function $f$ can be locally expanded as $f_{ij} = f(0) + a {e}_{ij} \cdot \nabla f(0) + \mathcal{O}(a^2)$,
where $a$ is the lattice spacing and ${e}_{ij}$ is the unit vector pointing in the correct direction (we are generically considering $d$ dimensional lattices). We can similarly expand $n_j$ but need to go to an additional order for $\hat{n}_j$ in to recover the correct expansion
\begin{equation}
    \e^{\hat{n}_j - \hat{n}_i} - 1 = a {e}_{ij} \cdot \nabla \hat{n}_i + \frac{a^2}{2} \big[ ({e}_{ij} \cdot \nabla)^2 \hat{n}_i + ({e}_{ij} \cdot \nabla \hat{n}_i)^2\big].
\end{equation}
Combining all terms and expanding up to order $a^2$, the sum over neighbouring sites in the action gives
\begin{equation}
\begin{aligned}
    \sum_{(i,j)} n_i W_{ij}(\e^{\hat{n}_j - \hat{n}_i} - 1) &=  \sum_{(i,j)}\Big[ a f(0) n_i(1-n_i) {e}_{ij} \cdot \nabla \hat{n}_i + \frac{a^2}{2} f(0) n_i(1-n_i) ({e}_{ij} \cdot \nabla)^2 \hat{n}_i\\
    &\qquad + \frac{a^2}{2} f(0) n_i(1-n_i) ({e}_{ij} \cdot \nabla \hat{n}_i)^2 - a^2 f(0) n_i ({e}_{ij} \cdot \nabla n_i)({e}_{ij} \cdot \nabla \hat{n}_i)\\
    &\qquad + a^2 n_i(1-n_i) ({e}_{ij} \cdot \nabla f(0))({e}_{ij} \cdot \nabla \hat{n}_i) \Big].
\end{aligned}
\end{equation}
The first term of order $a$ vanishes by symmetry, as the unit vector retains its sign and adjacent sites thus cancel out as $a \to 0$. In this continuous limit, the sum becomes an integral (time may be rescaled as required), and the action becomes
\begin{equation}
\begin{aligned}
    S[\{\rho,\hat{\rho}\}] = -\int \dd t \, \int \dd {x} &\, \bigg\{ -\hat{\rho} \partial_t \rho + f(0) \rho(1-\rho) \nabla^2 \hat{\rho} + f(0) \rho(1-\rho) (\nabla \hat{\rho})^2 - 2f(0) \rho \nabla \rho \cdot \nabla \hat{\rho} + 2\rho(1-\rho) \nabla f(0) \cdot \nabla \hat{\rho} \bigg\},
\end{aligned}
\end{equation}
where the factor 2 comes from the fact that summing squared terms over all neighboring sites $j$ always gives pairs of terms (for example on a $d=1$ lattice we have a derivative going forward and backward that are both evaluated at $i$ and so give twice an identical term). At this stage, one can finally perform several integration by parts to recover only terms factorized by $\hat{\rho}$ of $(\nabla \hat{\rho})^2$, corresponding to the deterministic and fluctuating contributions of the final action respectively.

Skipping details of this straightforward procedure, the generating functional is finally given by
\begin{equation}
\begin{aligned}
    S[\{\rho,\hat{\rho}\}] = -\int \dd t \, \int \dd {x} \,\bigg\{ -\hat{\rho} \partial_t \rho + f(0) \hat{\rho} \nabla^2 \rho - 2 \hat{\rho} \nabla \cdot (\rho(1-\rho) \nabla f(0)) + f(0) \rho (1-\rho)(\nabla \hat{\rho})^2\bigg\},
\end{aligned}
\end{equation}
corresponding to the Langevin equation
\begin{equation}
    \partial_t \rho = \nabla \cdot \Big[ f(0) \nabla \rho - 2\rho(1-\rho) \nabla f(0) + \sqrt{\rho (1-\rho)} \xi \Big]
\end{equation}
with the Gaussian white noise $\xi$ correlated as
\begin{equation}
    \langle \xi({x},t) \xi({x}',t') \rangle = 2 f(0) \delta^d({x}-{x}') \delta(t-t').
\end{equation}
The equations for the main text can then be recovered by taking the mobility $M[\rho] = \rho(1-\rho)$ and the classic logit decision rule leading to the transition rate
\begin{equation}
    f(\Delta\upsilon) = \frac{1}{1 + \e^{-\frac{\Delta \upsilon}{T}}},
\end{equation}
where $\Delta \upsilon$ is the change of utility proposed to the agent, which will be different whether they are an altruist or an individualist, and rescaling time appropriately. As noted in \cite{zakine2024socioeconomic}, this choice is non-unique however, as any function with identical value and first derivative in zero leads to the same field theory.}

\subsection{Linear stability analysis}
\label{app:LSA}
\subsubsection{Two-population model}
We start from the coupled evolution equations
\begin{align}
    \partial_t \rhoA &= \partial_x\left(\rhoA(1-\rhoA-\rhoI)\partial_x \frac{\delta \mathcal{F}}{\delta \rhoA(x)}\right)\\
    \partial_t \rhoI &= \partial_x\left(\rhoI(1-\rhoA-\rhoI)\partial_x\mu_\mathrm{I}\right).
\end{align}
Taking $\rhoA(x,t) = \alpha \rho_0 + \epsilon \tilde{\rho}_\mathrm{A}(x,t)$, $\rhoI(x,t) = (1-\alpha) \rho_0 + \epsilon \tilde{\rho}_\mathrm{I}(x,t)$ and expanding the equations up to order $\epsilon$, we have the evolution of the perturbations at the linear order
\begin{align}
    \partial_t \tilde{\rho}_{\mathrm{A}} &= T(1 - (1-\alpha) \rho_0)\partial_x^2 \tilde{\rho}_{\mathrm{A}} + \alpha T\rho_0 \partial_x ^2 \tilde{\rho}_{\mathrm{I}} -  \alpha \rho_0 (1- \rho_0)\partial^2_x [ 2\tilde{\phi} u'(\rho_0)+\rho_0 u''(\rho_0) G*\tilde{\phi}]\\
    \partial_t \tilde{\rho}_{\mathrm{I}} &= T(1-\alpha\rho_0)\partial_x^2 \tilde{\rho}_{\mathrm{I}} + (1-\alpha) T \rho_0 \partial_x ^2 \tilde{\rho}_{\mathrm{A}} -  (1-\alpha) \rho_0 (1- \rho_0)\partial^2_x [ \tilde{\phi} u'(\rho_0)],
\end{align}
with $\tilde{\phi} = G*(\tilde{\rho}_\mathrm{A} + \tilde
{\rho}_{\mathrm{I}})$.
In Fourier space, these equations yield the linear system
\begin{equation}
    \partial_t \begin{bmatrix}
        \hat{\rho}_{\mathrm{A}}(k,t)\\
        \hat{\rho}_{\mathrm{I}}(k,t)
    \end{bmatrix}
    = K \begin{bmatrix}
        \hat{\rho}_{\mathrm{A}}(k,t)\\
        \hat{\rho}_{\mathrm{I}}(k,t)
    \end{bmatrix},
\end{equation}
with the stability matrix $K=$
\begin{align}
    -k^2T
    \begin{pmatrix}
      (1-(1-\alpha)\rho_0) - \alpha \rho_0 (1-\rho_0)\hat G(k)[2u'(\rho_0) +\rho_0 u''(\rho_0)\hat G(k)] 
      & \alpha  \rho_0 - \alpha \rho_0(1-\rho_0)\hat G(k)[2u'(\rho_0) +\rho_0 u''(\rho_0)\hat G(k)] \\
       (1-\alpha)\rho_0 -(1-\alpha)\rho_0(1-\rho_0)\hat G(k) u'(\rho_0)  
      &  (1-\alpha\rho_0) -(1-\alpha)\rho_0(1-\rho_0)\hat G(k) u'(\rho_0)
    \end{pmatrix}.
\end{align}
The eigenvalues can be computed explicitly. They read
\begin{align}
    \lambda_1 &=-k^2 T (1-\rho_0),\\
    \lambda_2 &= -k^2 \big(T - \rho_0(1-\rho_0) \hat{G}(k) \big[ (1+\alpha)u'(\rho_0)+ \alpha \rho_0 u''(\rho_0) \hat{G}(k) \big] \big),
\end{align}
which are always real, and thus seem to indicate that chasing cannot be observed from the homogeneous state (contrary to some cases when combining two individualistic populations with competing goals, as documented in \cite{zakine2024socioeconomic}). Clearly, $\lambda_1$ will always be negative, and we must therefore consider $\lambda_\mathrm{max} = \lambda_2$ to determine the spinodal.

\subsubsection{Single population simplification}
As mentioned in the main text, it is rather straightforward to convince oneself that the linear stability analysis about the \textit{homogeneous} state is the same in the simplification we propose as in our original model. To check that this is indeed the case, we perform the single population computation here. 
Starting now from Eq.~\eqref{eq:model2} of the main text, expanding it in the vicinity of the homogeneous state $\rho(x,t) = \rho_0 + \epsilon \tilde{\rho}(x,t)$ and writing the time evolution of the perturbation in Fourier space, we have
\begin{equation}
    \partial_t \hat{\rho}(k,t) = -k^2 \big(T - \rho_0(1-\rho_0) \hat{G}(k) \big[ (1+\alpha)u'(\rho_0) + \alpha \rho_0 u''(\rho_0) \hat{G}(k) \big] \big) \hat{\rho}(k,t).
\end{equation}
We immediately recognize that the criterion for linear stability is the same as above.

\subsection{Agent-based simulations}
\label{app:ABM}

The main text has focused on the coarsed-grained hydrodynamic description of our models. The locally conserved dynamics assumed that agents evolve locally in space, jumping from one site to some neighboring one. In order to ensure that the central conclusions of our study are robust to variations of this setting, as well as to lie closer to standard socioeconomic modeling, we consider a discrete version or our model with non-local moves, which appear more realistic from a practical standpoint. 

At every time step in the simulation, an agent (altruistic or individualistic) is randomly selected from an occupied site on the lattice, and is proposed to move to a randomly selected empty site. The difference in the agent's utility $\Delta \upsilon$ is computed based on the type of agent selected (either global or local), and the move is accepted with probability
\begin{equation}
    P(\Delta \upsilon) = \frac{1}{1+\e^{-\frac{\Delta \upsilon}{T}}},
\end{equation}
ensuring that detailed balance is satisfied in the case of a purely altruistic population. As shown in \cite{zakine2024socioeconomic}, detailed balance is violated for individualistic agents whenever the utility function is a nonlinear function of the \textit{local} perceived density $\phi$, consistent with the non-relaxational nature of the dynamics for $\alpha < 1$. Here, the perceived density is computed by performing a discrete convolution between the kernel $G$ and a binary occupation variable $n_i$, equal to one if site $i$ is occupied and zero if it is empty. The global utility for a system of size $\Omega$ is now defined as
\begin{equation}
    U = \sum_{i=1}^{\Omega} n_i u_i,
\end{equation}
with $u_i$ the (hypothetical) individual utility of an agent located at the associated site.

To measure the coexistence densities shown in Fig.~\ref{fig:binodals}(a) of the main text we initialize two-dimensional systems of width $L = 400$ and height $\ell = 100$ in a phase separated state, forming a slab with a concentrated region in the center. The system is left to evolve and the densities are measured by time-averaging the occupation variables $n_i$ once the system has reached a steady-state. In this experiment, we take $G$ to be a Gaussian kernel with characteristic width $\sigma = 7$, explaining the slight disparities with the one-dimensional noiseless PDE resolutions that may be performed on larger systems lying closer to the true mean-field limit $L \to \infty$, $\sigma \to \infty$ with $\sigma/L \to 0$.

As expected from the results of \cite{zakine2024socioeconomic} for the $\alpha = 0$ case, we find a very good agreement between the local hydrodynamics and the non-local simulations.

\subsection{Generalized thermodynamic mapping}

\subsubsection{Gradient expansion}
The ``well-mixed'' version of the model considers a single type of agents who interpolates between individualistic and altruistic behavior. The chemical potential thus interpolates between the individualistic chemical potential and the altruistic one.  The mean-field equation of motion reads
\begin{align}
    \partial_t \rho = \nabla\cdot[\rho(1-\rho)\nabla\mu_\mathrm{wm}],
\end{align}
with
\begin{align}
    \mu_\mathrm{wm} =  T \log\left(\frac{\rho}{1-\rho}\right) -u(\phi) -\alpha \int \rho(y) u'(\phi(y))G(x-y) \dd y .
\end{align}
To identify a good change of variable that yields the binodal densities, one expands the chemical potential, and one retains the leading order gradient terms. On thus obtains 
\begin{equation}
\begin{aligned}
    \mu_\mathrm{wm}([\rho],x) 
    = \mu_0(\rho) + \lambda(\rho) (\nabla \rho)^2 - \kappa(\rho) \nabla^2 \rho + O(\nabla^4).
    \label{eq:mu_gradient_appendix}
\end{aligned}
\end{equation}
The expansion is generic and can be made explicit for any smoothing kernel $G$. For simplicity, we assume $G$ to be Gaussian with variance $\sigma^2$, and we identify the different terms:
\begin{align}
    \mu_0(\rho) &=-u(\rho)-\alpha\rho u'(\rho) +T\log\left(\frac{\rho}{1-\rho}\right),\\
    \kappa(\rho) &=\frac{\sigma^2}{2}[(1+\alpha)u'(\rho) + 2\alpha \rho u''(\rho)],
    \label{eq:kappa_rho_sigma}\\
    \lambda(\rho) &= -\frac{\sigma^2}{2}\alpha[2u''(\rho)+\rho u'''(\rho)].
\end{align}

\subsubsection{Change of variable}

\label{app:Gen_thermo}
To find the binodal densities, one first defines the bijective change of variable $R(\rho)$ that satisfies $\kappa(\rho)R''(\rho)=-[\kappa'(\rho)+2\lambda(\rho)]R'(\rho)$, which here reads
\begin{equation}
    R''(\rho) = -\frac{(1-\alpha)u''(\rho)}{(1+\alpha)u'(\rho)+2\alpha\rho u''(\rho)} R'(\rho).
    \label{eq:change_of_var}
\end{equation}
Taking our utility function $u(\rho)=-|\rho-\rho^\star|^\gamma$, it turns out we always have
\begin{align}
    \frac{u'}{u''}=\frac{\rho^\star-\rho}{1-\gamma},
\end{align}
which simplifies the differential equation into
\begin{align}
    R'' = -\frac{(1-\alpha)(1-\gamma)}{(1+\alpha)(\rho^\star-\rho)+2\alpha(1-\gamma)\rho}R'.
\end{align}
This equation has to be solved on two distinct domains before ``gluing''.
The extremal density is 
\begin{align}
    \rho_m = \frac{1}{1+2\frac{\alpha}{\alpha+1}(\gamma-1)}\rho^\star,
\end{align}
and the solution for $R(\rho)$ reads
\begin{align}
    R(\rho)=C_1 (\rho-\rho_m)^\xi \,\Theta[\rho-\rho_m] + C_2 (\rho_m-\rho)^\xi\, \Theta[\rho_m-\rho],
\end{align}
with $\xi=1+\dfrac{(\alpha-1)(\gamma-1)}{1+\alpha(2\gamma-1)}$, and the constants $C_1$ and $C_2$ have to be adjusted such that $R$ is bijective and differentiable. Interestingly enough, we recover a linear change of variable, i.e. $\xi=1$ if and only if $\gamma=1$ (linear utility) or $\alpha=1$ (global utility). We can always shift $R$ with some constant also.
In the other way round, we have:
\begin{align}
    \rho = \rho_m + \sign(R) |R|^\frac{1}{\xi}
    \label{eq:R_to_rho}
\end{align}
and now the chemical potential
\begin{align}
\begin{split}
    \mu_0(R) =& -u(\rho(R))-\alpha\rho(R) u'(\rho(R)) +T\log(\rho(R)/(1-\rho(R)))\\
    =& \left|\rho_m + \sign(R)|R|^\frac{1}{\xi}-\rho^\star\right|^\gamma + \alpha \gamma (\rho_m +\sign(R)|R|^\frac{1}{\xi}) \sign\left(\rho_m + \sign(R) |R|^\frac{1}{\xi}-\rho^\star\right) \left|\rho_m + \sign(R)|R|^\frac{1}{\xi}-\rho^\star\right|^{\gamma-1} \\
    &+ T \log\left( \frac{\rho_m + \sign(R)|R|^{\frac{1}{\xi}}}{1-\rho_m - \sign(R)|R|^{\frac{1}{\xi}}}\right).
\end{split}
\end{align}
Note that this expression is completely independent of $\sigma$, as expected in the mean-field limit. For given values of the parameters $\gamma$, $\rho^\star$, $T$ and $\alpha$, the Maxwell construction or the double-tangent construction can be performed numerically, yielding the ``liquid'' and ``gas'' values of $R$, which then yields the coexistence densities with Eq.~\eqref{eq:R_to_rho}.

\subsubsection{Surface tension and nucleation}

The surface tension is computed from the generalized free energy density and the coexistence densities $R_g$ and $R_\ell$ and reads, see \cite{solon2018generalized1,solon2018generalized2},
\begin{align}
\zeta =\int_{R_g}^{R_\ell}\sqrt{2\kappa(R)\Delta g(R)}\dd R,
\label{eq:surface_tension_used}
\end{align}
with $\Delta g= \tilde g(\rho)-\tilde g(R_\ell)$ and $\tilde g(R)=g(R)-\mu_0(R_\ell) R$, the free energy density tilted by the chemical potential at phase coexistence. As such, the surface tension $\zeta$ depends on the interaction range $\sigma$ via $\kappa(\rho)$, see Eq.~\eqref{eq:kappa_rho_sigma}, that ultimately controls the width of the interface between the liquid and the gas. We thus normalize the tension by the length $\sigma$ in the main text.

To assess the nucleation rate, we need the quasi-potential at the critical radius $R_c$. Following Ref.\cite{cates2023classical}, the critical radius in $d=2$ is given by
\begin{align}
    R_c = \frac{\zeta}{\epsilon \Delta R \mu_0'(\rho_g)},
\end{align}
with $\epsilon=\rho_0-\rho_g$ the saturation of the homogeneous state in the binodal and with $\Delta R=R_\ell-R_g$.
Finally, the quasipotential at the critical radius of a liquid droplet surrounded by the gas reads
\begin{align}
 V(R_c) = \pi \frac{\rho_\ell -\rho_g}{R_\ell-R_g} \frac{\zeta^2}{\epsilon (R_\ell-R_g) \mu_0'(\rho_g)}.
\end{align}
Equivalently, the quasipotential to nucleate a bubble of gas in a liquid is obtained by exchanging the indices $\ell\leftrightarrow g$ in the formula above.
In practice, since all quantities are $\alpha$ dependent, the effect of $\alpha$ on the quasipotential is not easily apprehensible.

\begin{figure}
    \centering
    \includegraphics[width=.5\linewidth]{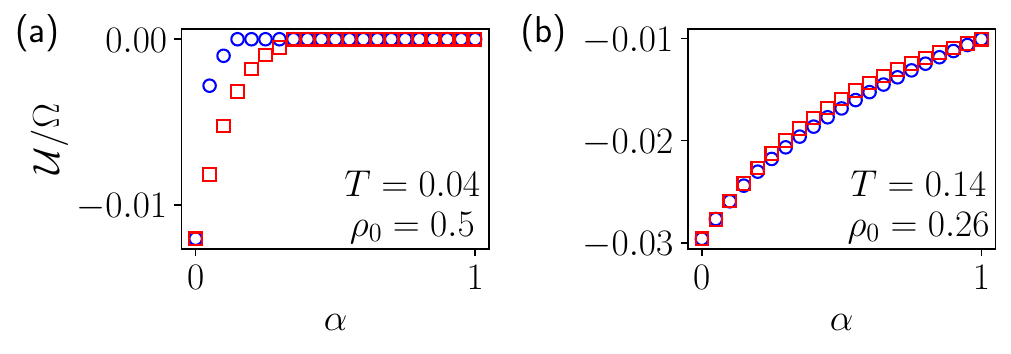}
    \caption{Evolution of the global utility $\mathcal{U}[\rho] = \int \dd x \, u(\phi([\rho],x)) \rho(x)$ in the steady state obtained from the numerical resolution of the noiseless PDE ($L=1000$, $\sigma = 10$) with the level of altruism $\alpha$ for two populations ($\circ$) and the single population simplification ($\square$). (a) Low temperature where there is a significant difference between the two prescriptions. (b) Higher temperature, where the well-mixed approximation holds.}
    \label{fig:global_utility}
\end{figure}

{
\subsection{Comparison between the single and two-population model global utilities}
As stated in the main text, we can directly compute the global utility from coexistence densities in the thermodynamic limit,
\begin{equation}
    \frac{\mathcal{U}}{L^d} = \frac{\rho_0 - \rho_g}{\rho_\ell - \rho_g} u(\rho_\ell) + \frac{\rho_\ell - \rho_0}{\rho_\ell - \rho_g} u(\rho_g) + O\left(\frac1L\right).
\end{equation}
Given the difference between the binodal curves of the original two-population and the simplified single population models (Fig.~\ref{fig:binodals}(a) vs (b)) we expect that they will lead to different global utilities in the steady state at sufficiently low temperatures and intermediate values of $\alpha$. We verify and quantify this effect in Fig.~\ref{fig:global_utility}. As visible in Fig.~\ref{fig:global_utility}(a), the effect of altruism is much less effective in the single population case. At higher temperatures, we recover that the models are very nearly equivalent and that there is therefore little to no difference in the global utility, see Fig.~\ref{fig:global_utility}(b).}

